# Self-started unidirectional operation of a fiber ring soliton laser without an isolator

**L. M. Zhao, D. Y. Tang, and T. H. Cheng**

School of Electrical and Electronic Engineering, Nanyang Technological University, Singapore

**C. Lu**

Department of Electronic and Information Engineering, Hong Kong Polytechnic University, Hung Hom, Kowloon, Hong Kong, China

We demonstrate self-started mode-locking in an Erbium-doped fiber ring laser by using the nonlinear polarization rotation mode-locking technique but without an isolator in cavity. We show that due to the intrinsic effective nonlinearity discrimination of the mode-locked pulse propagating along different cavity directions, the soliton operation of the laser is always unidirectional, and its features have no difference to that of the unidirectional lasers with an isolator in cavity.

Self-started, passively mode-locked Erbium-doped soliton fiber lasers as an attractive ultrashort pulse source at the 1.55 μm wavelength have been extensively investigated [1-3]. Among the different passive mode-locking techniques, the nonlinear polarization rotation (NPR) method has been widely used due to its simplicity and easy of implementation [2-4]. Conventionally, fiber lasers mode-locked with the NPR method adopt a unidirectional ring cavity. It was shown that such a laser cavity configuration could reduce the spurious cavity reflections and decrease the self-started mode-locking threshold. Indeed, as shown by Tamura et al. based on a mode-locked fiber laser and a solid-state laser, significant threshold power deduction could be achieved with a unidirectional ring cavity than a linear cavity [2]. In the early days of the fiber laser development due to the lack of high power pump source, reduction of the laser mode-locking threshold is necessary. In this paper we show experimentally that one could also obtain self-started mode locking and soliton operation in the bidirectional fiber ring lasers. Under continuous wave operation a bidirectional fiber laser can lase simultaneously on both directions with equal output intensity. However, after mode-locking, due to the cavity nonlinearity induced directional symmetry breaking the laser will always operate unidirectional. One could actually predefine the operation direction of the laser through the cavity design. We further show experimentally that the soliton operation of a bidirectional fiber ring laser has no measurable difference to those of the unidirectional soliton fiber lasers.

The fiber laser we used has a cavity configuration as illustrated in Fig. 1. The cavity comprises a segment of 2.7-meter-long Erbium-doped fiber (EDF-1480-T6) with group velocity dispersion (GVD) parameter of 10 (ps/nm)/km, a segment of 6-meter-long dispersion compensation fiber (DCF) with GVD parameter of -0.196 (ps/nm)/km, and all other fiber segments are standard single mode fiber (SMF). The NPR technique is used to achieve the self-started mode locking in the laser. Either the unidirectional operation or the bidirectional operation of the laser was studied. For the unidirectional operation of the laser, a polarization-independent isolator and a polarizer were inserted in the cavity to force the unidirectional operation; while for the bidirectional operation, the polarization-independent isolator was removed. Two polarization controllers, one consisting of two quarter-wave plates and the other one quarter-wave plate, were used to control the polarization of the light. The polarization controllers, the polarization-independent isolator and the polarizer were mounted on a 7-cm-long fiber bench. The laser was pumped by a high power Fiber Raman Laser source (BWC-FL-1480-1) of wavelength 1480 nm. The output was outlet from a 10% output coupler. The WDM and coupler are both made of standard SMF.

We first investigated the unidirectional operation of the laser with the polarization-independent isolator in cavity. The laser is a typical conventional soliton fiber laser mode-locked by the NPR technique. Self-started mode locking is easily obtained in the laser provided the orientations of the polarization controllers are appropriately set. Typical soliton spectra with the characteristic sidebands are shown in Fig. 2. Multiple soliton operation and soliton hysteresis are observed in the laser, which are identical as those reported previously [5, 6]. With an isolator in cavity the laser operation direction is determined by the isolator direction. In Fig. 2 we also compare the soliton spectra of the laser under the two different operation directions. While when the laser operates in the clockwise direction, the residual 1480nm pump is clear to see on the spectrum, it is blocked by the isolator when the laser operates in the anticlockwise direction. The measured soliton spectra have obviously different profiles, indicating that the effective nonlinearity of the soliton pulse experienced along the different cavity directions is different. Under exactly the same pump strength, e.g. 50.5 mW, the average output power of the laser under different operation directions is different.

The cavity transmission of fiber lasers mode-locked with the NPR technique is equivalent to the light transmission through a polarizer and an analyzer with a weakly birefringent fiber between them, which can be described as [6]

$$T = \sin^2\theta\sin^2\varphi + \cos^2\theta\cos^2\varphi + \frac{1}{2}\sin 2\theta\sin 2\varphi\cos\left[\ \Delta\Phi_l + \Delta\Phi_{nl}\ \right] \quad (1)$$

whrere $\theta$ and $\varphi$ are the orientation angle of the polarizer and analyzer with respect to the fast axis of the weakly birefringent fiber, respectively, and $\Delta\Phi_l$ and $\Delta\Phi_{nl}$ are the phase delay between the two orthogonal polarization components of light caused by the linear and nonlinear fiber birefringence. In the practice of a fiber ring laser as shown in Fig.1, the polarizer in the cavity plays both the role of a polarizer and an analyzer for generating the self-started mode-locking. Obviously, the laser cavity transmission comprises two parts, a linear cavity transmission part determined by the cavity configuration and parameter setting. In the case of a bidirectional ring laser, this part is independent of the operation direction of the laser, a dynamical cavity transmission part, which is the cavity nonlinearity dependent. In the case of a bidirectional fiber ring laser the nonlinear phase delay is generally the light propagation direction dependent. We believe that it is this dynamical cavity loss difference that has resulted the observed soliton spectral difference and average mode-locked output power difference.

We then removed the isolator and investigated the bidirectional operation of the laser. Experimentally we found that self-started mode locking can still be achieved even the laser cavity is bidirectional. Nevertheless, the mode-locking threshold is significantly increased. While with the unidirectional cavity the self-started mode-locking threshold is about 120mw, it increased to about 800mw with the bidirectional cavity. We further compared the laser operation before and after the mode-locking. While operating at the continuous wave state, the laser emits simultaneously bidirectional with nearly identical output intensity, once the laser is mode-locked it always operates unidirectional in the clockwise direction. It shows that under the mode-locked operation the bidirectional cavity is directionally asymmetric. There is a mode-locking priority direction. Analyzing our laser cavity configuration, along the clockwise direction, after being amplified the mode-locked pulse first propagates in the DCF and then traverses the cavity polarizer. Due to that the DCF has only weak dispersion, large nonlinear phase shift is generated. While if the pulse propagates along the anticlockwise direction, after being amplified, it first encounters the cavity output and then the polarizer, finally propagates in the DCF. As when the pulse propagates in the DCF it has only weak pulse energy, weak nonlinearity would be generated. The mode locking mechanism of the laser is the NPR. Cavity with larger nonlinearity would be easier to be mode locked. Therefore, we attribute the existence of the mode-locking priority direction as a result caused by the different nonlinearity experienced by the mode-locked pulse along the different directions. It is the cavity nonlinearity asymmetry that determines the mode-locked laser operation direction. In our laser in order to enhance the cavity nonlinearity discrimination and reduce the mode-locking threshold, we have deliberately added a piece of the DCF in the cavity. In the practice due to the different order of the cavity components along different cavity directions etc a bidirectional ring laser can easily become dynamically asymmetric. Therefore, it is expected that this effect could be a general feature of the mode-locked bidirectional soliton ring lasers.

Once mode-locked, the soliton operation of the bidirectional fiber laser shows no difference to those of the unidirectional lasers. Fig. 3a shows a typical soliton spectrum of the laser measured, which has clearly the characteristics of the soliton spectrum shown in Fig. 2a. The inset in Fig. 3a is the corresponding oscilloscope trace of the output soliton train. The repetition rate of the soliton train is about 17.8 MHz, which corresponds to the cavity length. Either the multiple soliton operation or the noiselike pulse emission [7, 8] was experimentally observed. Under the multiple soliton operation the solitons also automatically form the various operation modes reported previously [5,6], such as the soliton bunching, random stable multiple soliton distribution and stochastic soliton movements.

Any optical isolators have unavoidably insertion loss. An obvious advantage of the laser compared with the unidirectional lasers is that larger average laser output power could be obtained. However, without isolator in cavity there is also a serious drawback. Experimentally it was observed that associated with the soliton emission in the clockwise direction, a weak soliton-like emission is also detectable in the anticlockwise direction. In Fig. 3b we have shown the optical spectrum measured simultaneously in the anticlockwise direction. The emission strength in the direction is about 11dB weaker than that of the clockwise direction. Furthermore, it has almost the same spectral profile as that of the laser emission. Careful examination on the oscilloscope traces measured along the two different directions shows that both have the same pulse pattern, suggesting that the weak counter-propagation direction emission in the laser is only a weak image of the laser emission. It is caused by the cavity backscattering (or reflection) effect.

In conclusion, we have shown experimentally that self-started mode locking can still be achieved in the bidirectional fiber ring lasers with the NPR technique. However, due to the intrinsic cavity nonlinearity asymmetry along the two propagation directions, the mode-locked bidirectional ring laser still operate unidirectional, and its soliton operation has no difference to those of the unidirectional lasers.

## Figure Caption:

Fig. 1 Experimental setup. $\lambda/4$ : quarter-wave plate; I: polarization-independent isolator; P: polarizer; WDM: wavelength-division multiplexer; EDF: Erbium doped fiber; DCF: dispersion compensation fiber.

Fig. 2 typical soliton spectra of the laser. (a) Laser operates in the clockwise direction; (b) laser operates in the anticlockwise direction.

Fig. 3 soliton spectra of the bidirectional laser. (a) clockwise direction; (b) anticlockwise direction.

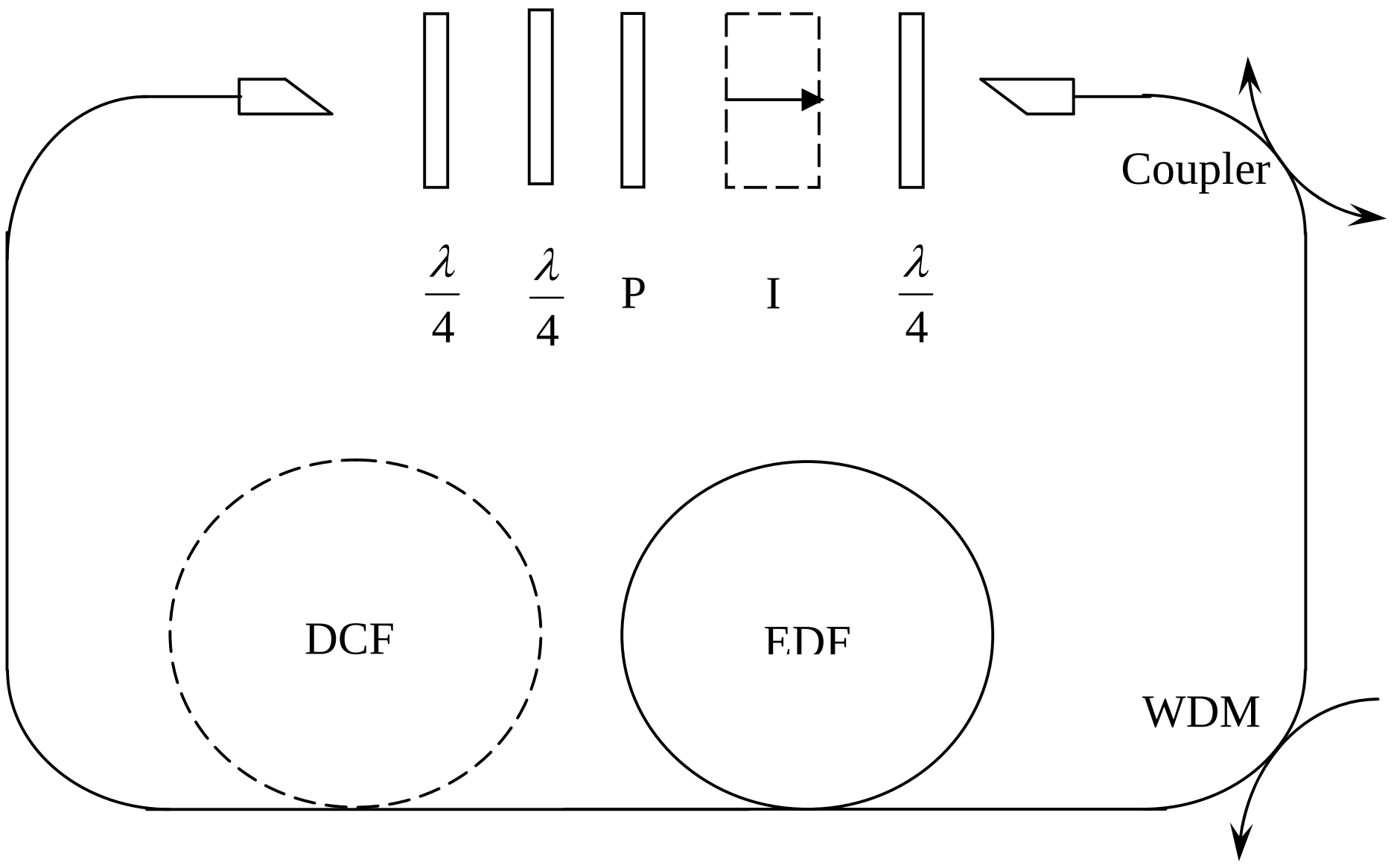

Fig. 1 L. M. Zhao et al. "Self-started passively mode-locked bidirectional soliton fiber ring laser"

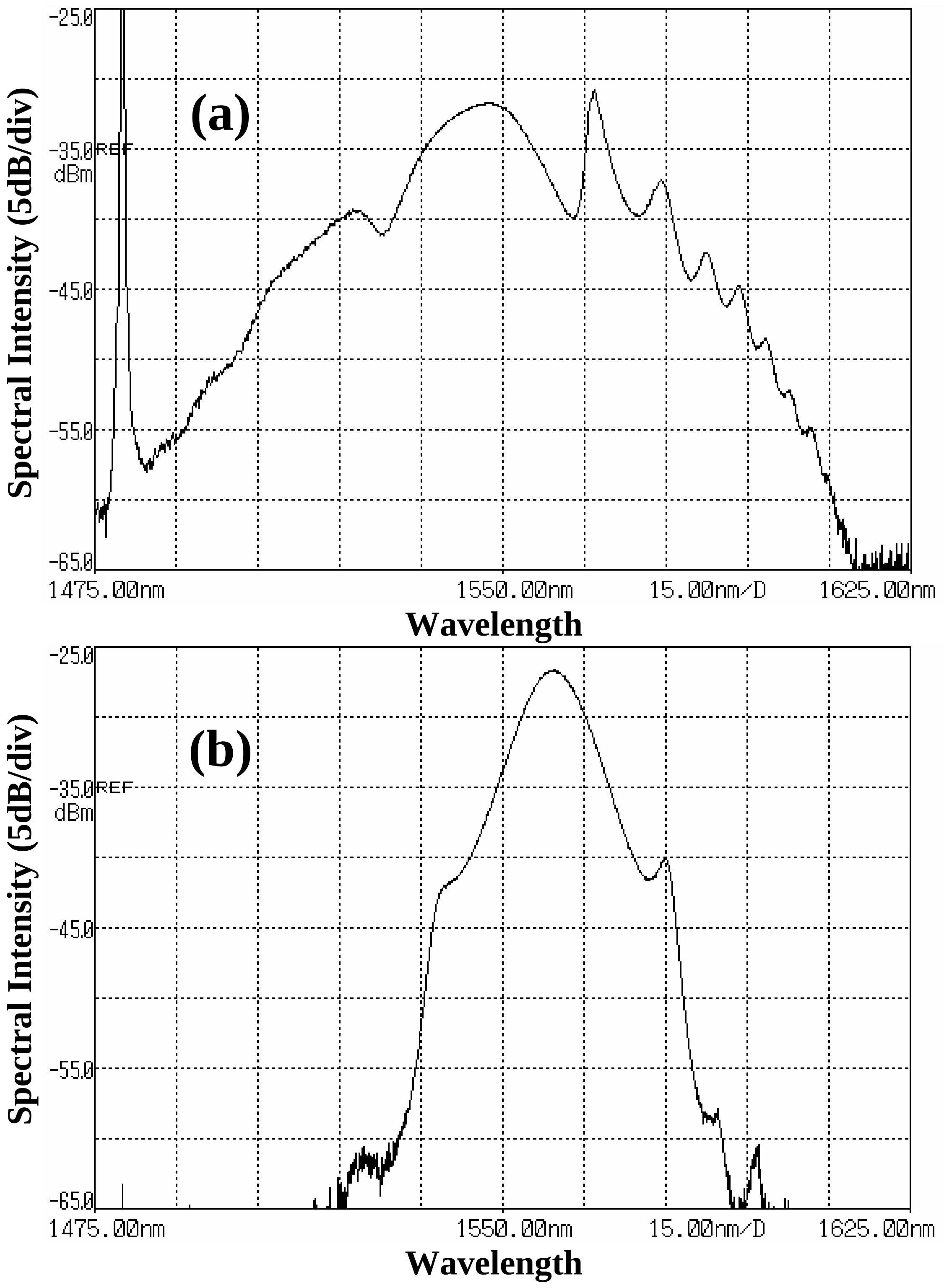

Fig. 2 L. M. Zhao et al. "Self-started passively mode-locked bidirectional soliton fiber ring laser"

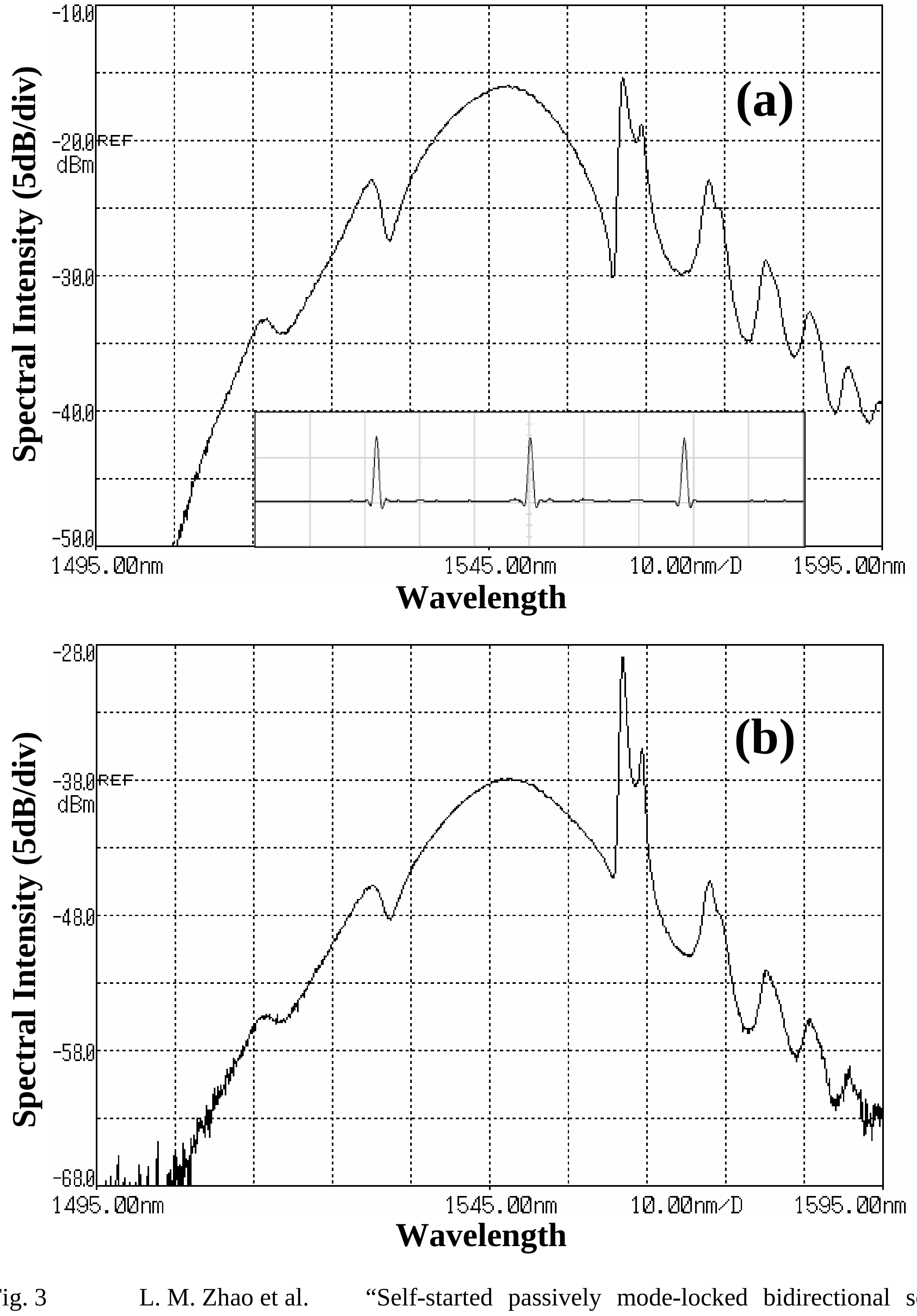

Fig. 3 L. M. Zhao et al. “Self-started passively mode-locked bidirectional soliton fiber ring laser”